\documentclass[12pt]{article}

\usepackage{amsthm,amsmath,amsfonts,amssymb}
\usepackage{graphicx}

\usepackage{natbib}
\usepackage{times}
\usepackage{bm}
\usepackage{bbm}
\usepackage{tikz}
\usepackage{color}
\usepackage{multirow}
\usepackage{float}
\usepackage{enumitem}
\usepackage{booktabs,caption,subcaption}
\usepackage{titlesec}
\usepackage{diagbox}
\usepackage{xr}

\usepackage[margin=1in]{geometry}
\allowdisplaybreaks
\usepackage{setspace}
\newcommand{\bJ}{\boldsymbol{J}}
\newcommand{\bj}{\boldsymbol{j}}

\newcommand{\bU}{\boldsymbol{U}}

\newcommand{\bV}{\boldsymbol{V}}

\newcommand{\bX}{\boldsymbol{X}}
\newcommand{\bx}{\boldsymbol{x}}

\newcommand{\by}{\boldsymbol{y}}

\newcommand{\bbeta}{\boldsymbol{\beta}}

\makeatletter
\DeclareFontFamily{U}  {MnSymbolF}{}
\DeclareSymbolFont{symbolsMN}{U}{MnSymbolF}{m}{n}
\SetSymbolFont{symbolsMN}{bold}{U}{MnSymbolF}{b}{n}
\DeclareFontShape{U}{MnSymbolF}{m}{n}{
    <-6>  MnSymbolF5
   <6-7>  MnSymbolF6
   <7-8>  MnSymbolF7
   <8-9>  MnSymbolF8
   <9-10> MnSymbolF9
  <10-12> MnSymbolF10
  <12->   MnSymbolF12}{}
\DeclareFontShape{U}{MnSymbolF}{b}{n}{
    <-6>  MnSymbolF-Bold5
   <6-7>  MnSymbolF-Bold6
   <7-8>  MnSymbolF-Bold7
   <8-9>  MnSymbolF-Bold8
   <9-10> MnSymbolF-Bold9
  <10-12> MnSymbolF-Bold10
  <12->   MnSymbolF-Bold12}{}
\DeclareMathSymbol{\tbigtimes}{\mathop}{symbolsMN}{2}
\newcommand*{\bigtimes}{%
  \DOTSB
  \tbigtimes
  \slimits@ 
}
\makeatother

\begin{document}

\begin{singlespace}

\title{\bf Parallel-and-stream accelerator for computationally fast supervised learning}

\author{Emily C. Hector \\Department of Statistics, North Carolina State University\\
Lan Luo \\ Department of Statistics and Actuarial Science, The University of Iowa\\
and\\
Peter X.-K. Song\\Department of Biostatistics, University of Michigan}

\date{}

\maketitle

\begin{abstract}
Two dominant distributed computing strategies have emerged to overcome the computational bottleneck of supervised learning with big data: parallel data processing in the MapReduce paradigm and serial data processing in the online streaming paradigm. Despite the two strategies' common divide-and-combine approach, they differ in how they aggregate information, leading to different trade-offs between statistical and computational performance. In this paper, we propose a new hybrid paradigm, termed a {\em Parallel-and-Stream Accelerator (PASA)}, that uses the strengths of both strategies for computationally fast and statistically efficient supervised learning. PASA's architecture nests online streaming processing into each distributed and parallelized data process in a MapReduce framework. PASA leverages the advantages and mitigates the disadvantages of both the MapReduce and online streaming approaches to deliver a more flexible paradigm satisfying practical computing needs. We study the analytic properties and computational complexity of PASA, and detail its implementation for two key statistical learning tasks. We illustrate its performance through simulations and a large-scale data example building a prediction model for online purchases from advertising data.
\end{abstract}

\noindent%

\end{singlespace}

\vfill

\newpage

\section{Introduction}
With the recent explosion in data collection, two core computational strategies have emerged for fast and scalable computing with Big Data: parallel data processing in the MapReduce paradigm and serial data processing in the online streaming paradigm. From a data analytics perspective, these two paradigms provide two distinct information aggregation principles, leading to different statistical properties and power to reach meaningful findings. The ultimate goal of improving computing speed is to facilitate the search for relevant data features and generate new knowledge through valid and efficient statistical learning approaches. We propose a new computational paradigm, the parallel-and-stream accelerator (PASA), that seamlessly integrates these two computational strategies to address a broad range of statistical learning problems, yielding a modern expedited process for transforming data into knowledge.\\
Supervised learning via regression analysis has played a central role in machine learning as a powerful tool for model-based classification, feature extraction, prediction and quantification of dose-response relationships. We consider the regression analysis of $N$ independent observations $\left\{ y_i, \bx_i \right\}_{i=1}^N$, $y_i$ denoting the outcome and $\bx_i \in \mathbb{R}^p$ explanatory variables, where the sample size $N$ may be so big that a direct analysis of the whole data using conventional methodology is computationally intensive or prohibitive. We model the relationship between the mean $\mathbb{E}(y_i)$ of outcome $y_i$ and explanatory variables $\bx_i$ through a generalized linear model (GLM): $\mathbb{E}(y_i\rvert \bx_i)=g(\bx^\top_i \bbeta)$, $i=1, \ldots, N$, where $g$ is a known link function and $\bbeta \in \mathbb{R}^p$ is the parameter vector of interest with a fixed dimension. The GLM class of models includes important nonlinear regression models, such as logistic regression, that are widely used in classification, inference and prediction settings with applications in public health and industry, among others.

The MapReduce strategy, due to its scalability, is one of the preferred strategies to handle the intensive computing demands of estimating $\bbeta$ when the sample size $N$ is very large. Essentially, this approach divides data into smaller data blocks and analyses individual blocks separately, typically in parallel. The results of these separate analyses are then combined using, for example, summary statistics \cite{Glass}, estimating functions \cite{Hansen} or $p$-value functions \cite{Xie-Singh-Strawderman}; see also \cite{Zellner, Jordan, Wang-Wang-Song-2012, Tang-Song} and references therein. The main appeal of this approach is a trade-off between statistical efficiency and computational speed achieved by selecting the number and size of data blocks. An approach with more and smaller blocks sacrifices statistical efficiency for lower computational cost, whereas an approach with fewer and larger blocks retains desirable statistical efficiency at the price of increased computational cost. For the latter approach with large data blocks, using a computationally inexpensive and statistically efficient estimation procedure in each large data block, such as renewable learning \cite{Luo-Song-2020}, would mitigate the trade-off between computational speed and statistical efficiency.

The renewable learning proposed by \cite{Luo-Song-2020} is an online estimation and inference method that uses the Rho architecture to maintain low computing cost with no loss of statistical efficiency. This streaming approach divides data into smaller data batches and updates parameter estimates sequentially from the first to the last data batch using only summary statistics and without re-accessing individual-level raw data from previous data batches. This sequential processing is similar in spirit to the stochastic gradient descent algorithm that has been extensively used in the field of online learning \cite{Robbins1951,Sakrison1965,Toulis2015MLE}. Statistical inference, however, has been largely ignored in most of the existing procedures. The renewable learning method proposed by~\cite{Luo-Song-2020} fills this gap. This approach appears more computationally expensive than divide-and-combine since it does not leverage parallel computing, but has no loss of statistical power. In contrast, divide-and-combine is computationally fast but results in a loss of statistical efficiency. The development of PASA is motivated by the desire to integrate these two approaches to achieve desirable computational and statistical properties.

\begin{figure}[ht]
\centering
\includegraphics[width=0.7\textwidth]{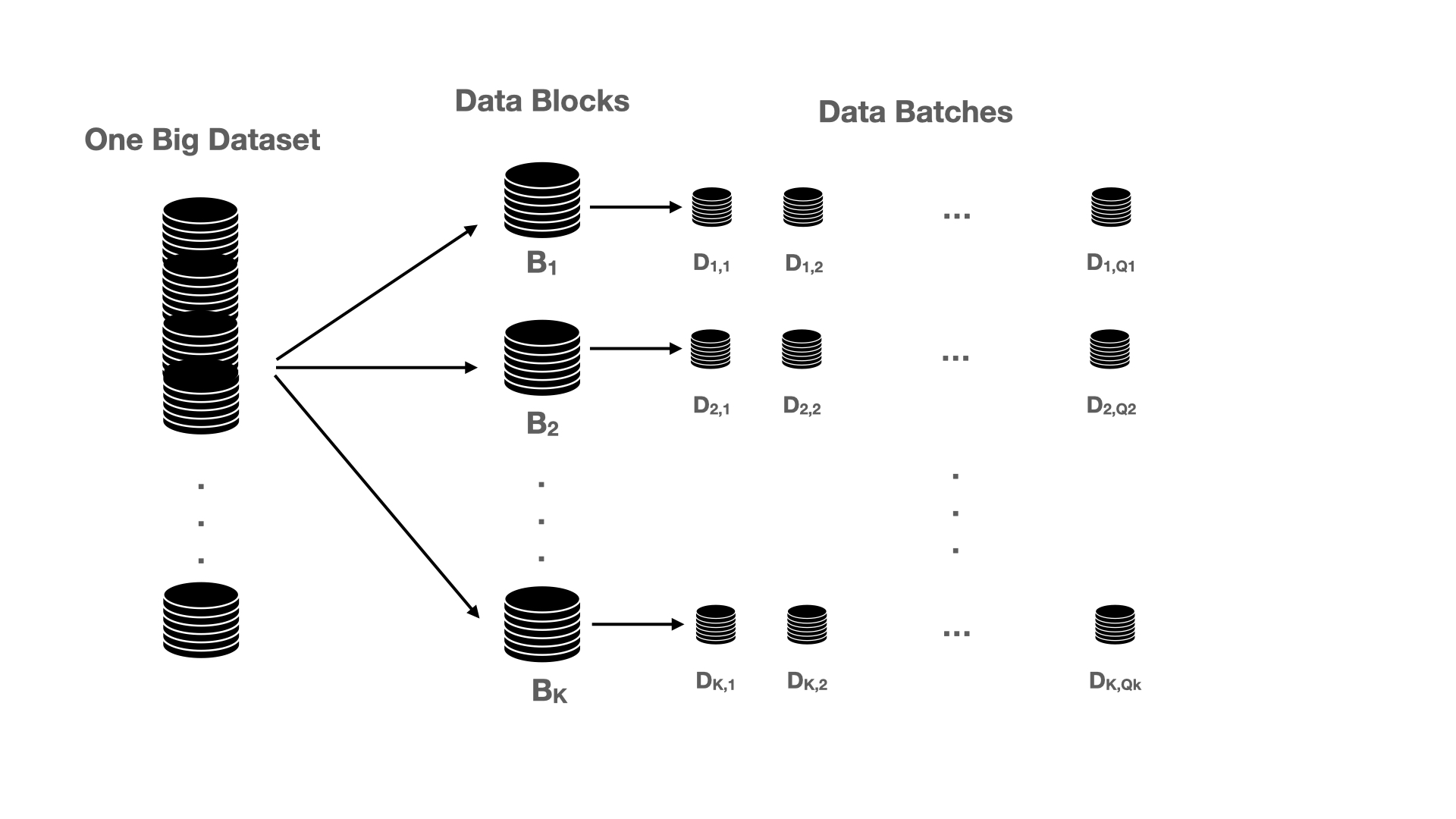}
\caption{Flowchart of PASA: A big dataset is first randomly partitioned into $K$ parallel data blocks, and each data block is then processed sequentially via renewable learning. \label{fig-dataflow}}
\end{figure}

While parallel and streaming approaches have been widely studied in the literature, to our knowledge they have never been combined. This is primarily because the technical challenges of statistical inference with big data have only recently been exposed. We propose the Parallel-and-Stream Accelerator (PASA), a hybrid approach that borrows strength from the two existing paradigms to overcome the computational bottleneck of statistical learning and inference in the GLM with big data. PASA divides a big dataset into large blocks and performs a streaming analysis in the Rho architecture within each block, as visualized in Figure \ref{fig-dataflow}. As a result, PASA benefits from the tremendous computational gains of parallel computing with a much lower order of bias from sequential computing to deliver a computationally and statistically efficient procedure. Clearly, the currently popular MapReduce and online approaches are special cases of the proposed hybrid method, which leverages the algorithmic power of both in a broad range of applications. PASA can be tailored to adapt to various data types and sizes through the specification of the link function $g$ and the number of data blocks and batches. 

The rest of this paper is organized as follows. Section \ref{sec:setup} describes the formal modeling setup and statistical framework. Section \ref{sec:alg} presents the proposed algorithm and corresponding asymptotic properties and computational complexity. Section \ref{sec:examples} details the proposed algorithm for two important statistical models, the linear and logistic regression models. Section \ref{sec:simulations} illustrates the performance of the proposed algorithm with simulations. Section \ref{sec:application} illustrates the implementation of PASA with the analysis of online advertising data. Section \ref{sec:discussion} concludes. Technical details are deferred to the Appendices.

\vspace{-2em}
\section{Setup}
\label{sec:setup}
Consider the regression setting with outcome $y$ drawn from a common parametric distribution conditional on $p$ explanatory variables $\bx =(x_1,\ldots, x_p)^\top$: $(y_i;\bx_i)\sim f(y;\bx,\bbeta_0,\phi_0)$, $i=1,\dots,N$ independently, where $\bbeta_0\in\mathbb{R}^p$ is the true value of the parameter of interest and $\phi_0$ is the true value of a nuisance parameter. Suppose the sample size $N$ is so big that a centralized analysis of the data is computationally prohibitive on one computing node. We consider the GLM setting in which the distribution $f(y; \bx, \bbeta, \phi)$ belongs to the family of exponential dispersion (ED) models~\cite{Jorgensen1997} whose associated log-likelihood function takes the form
\begin{align*}
\label{eq:ED_model}
\ell_{N}(\bbeta,\phi)
&=\sum_{i=1}^{N}\log f(y_i;\bx_i,\bbeta,\phi) = \sum_{i=1}^N \log a(y_i;\phi) - \frac{1}{2\phi}\sum_{i=1}^N d(y_i;\mu_i),
\end{align*}
where $d_i(y_i;\mu_i)$ is the unit deviance function involving the mean parameter $\mu_i=\mathbb{E}(y_i \rvert \bx_i)$, and $a(\cdot)$ is a suitable normalizing factor depending only on the dispersion parameter $\phi>0$. The dispersion $\phi$ corresponds to the variance parameter in the Normal distribution for the linear model, and is equal to 1 in the Bernoulli distribution for the logistic model. Given a specific distribution, the deviance function is known. For example, the deviance function of the Normal distribution is $d(y_i; \mu_i) = (y_i - \mu_i)^2$, the well-known least squared loss function. The deviance function for the Binomial distribution is $d(y_i;\mu_i) = 2[ y_i\log(y_i/\mu_i) + (1-y_i) \log\{ (1-y_i)/(1-\mu_i) \} ]$. The deviance function essentially serves as a nonlinear distance measure between data and model that is specified by a systematic component of the form $\mu_i=g(\bx_i^\top\bbeta)$. See more details in Chapter 2 of \cite{Song}.

Denote the unit score function for a single observation by $\bU(y_i;\bx_i,\bbeta):= \nabla_{\bbeta}d(y_i;\mu_i) = \partial d(y_i;\mu_i)/\partial \bbeta$. The offline maximum likelihood estimator (MLE) of the regression parameter $\bbeta$, denoted by $\widehat{\bbeta}$, is the root of the score equation:  $\sum_{i=1}^N \bU(y_i;\bx_i,\bbeta) = \bm{0}$. This offline MLE can be obtained numerically with the Newton-Raphson or Fisher-scoring algorithms, which have been implemented in many existing software packages such as SAS \texttt{PROC GENMOD} and the R function \texttt{glm}.\\
The computation of the offline MLE is contingent upon the central computing node having adequate computing capacity to iteratively search for the optimal parameter value. When this centralized computation is computationally prohibitive due to the large sample size, an appropriate algorithmic strategy is required. Our proposed PASA delivers computationally fast learning and inference in the GLM with minimal loss of statistical efficiency.

\section{Algorithm}
\label{sec:alg}

As visualized in Figure \ref{fig-dataflow}, PASA randomly splits the $N$ independent observations to form $K$ disjoint blocks, each denoted by $\mathcal{B}_k=\left\{\by_{k}, \bX_{k} \right\}$  with respective sample sizes $n_k=|\mathcal{B}_k|$, $k=1, \ldots, K$, such that $\sum_{k=1}^K n_k=N$. Due to independence between observations, the $K$ blocks are independent. The primary task is to conduct both estimation and inference for $\bbeta$ over the entire data. Given the data split, this task becomes a divide-and-combine procedure: the first step is to estimate $\bbeta$ and related inferential quantities in block $k$, and the second step is to integrate estimators from all blocks into one unified estimator and inferential procedure.

\subsection{Streaming algorithm}
\label{subsec:alg:stream}
In block $k \in \{1, \ldots, K\}$, we sequentially process the $n_k$ samples in a series of $Q_k$ data batches, denoted by $\mathcal{D}_{k,1} = \{\by_{k,1}, \bX_{k,1}\},\dots,\mathcal{D}_{k,Q_k}=\{\by_{k,Q_k},\bX_{k,Q_k} \}$, where $\by$ and $\bX$ are the generic notations of outcome vectors and associated covariate matrices. For convenience, slightly abusing the notation, we use $\mathcal{D}_{k,b}$ as the respective sets of indices for subjects in batch $\mathcal{D}_{k, b}$. Let $s_{k,b}=|\mathcal{D}_{k,b}|$ be the sample size in data batch $\mathcal{D}_{k,b}$, so that $\sum_{b=1}^{Q_k}s_{k,b}=n_k$. Note that in the streaming updating, data batch size $s_{k,b}$ is finite, and $n_k\to\infty$ is driven by $Q_k\to\infty$. Furthermore, we use $\bU_{k,b}(\mathcal{D}_{k,b};\bbeta) = \sum_{i\in\mathcal{D}_{k,b}}\bU(y_i;\bx_i,\bbeta)$ to denote the score function of data batch $\mathcal{D}_{k,b}$, and $\bJ_{k,b}(\mathcal{D}_{k,b};\bbeta) = - \nabla_{\bbeta} \bU_{k,b} (\mathcal{D}_{k,b}; \bbeta)$ is the corresponding negative Hessian matrix.

The renewable learning given in \cite{Luo-Song-2020} sequentially updates both point estimates and covariance matrices (or the Fisher Information) of $\bbeta$. It begins with the maximum likelihood estimator $(\widetilde{\bbeta}_{k,1}, \widetilde{\phi}_{k,1})$ of $(\bbeta, \phi)$ using the first data batch $\mathcal{D}_{k,1}$. For $j=2,\dots, Q_k$, instead of fitting models from scratch, previous estimators $( \widetilde{\bbeta}_{k,b-1}, \widetilde{\phi}_{k,b-1})$ are sequentially updated to $(\widetilde{\bbeta}_{k,b}, \widetilde{\phi}_{k,b})$ when the next data batch $\mathcal{D}_{k,b}$ is considered. After the updating, we store $\left\{\widetilde{\bbeta}_{k,b},\bJ_{k,b}(\mathcal{D}_{k,b};\widetilde{\bbeta}_{k,b}),\widetilde{\phi}_{k,b}\right\}$.

The updating of $\bbeta$ proceeds by solving the following incremental estimating equation:
\begin{equation}\label{eq:incremental_EE}
\sum_{j=1}^{b-1} \bJ_{k,j}(\mathcal{D}_{k,j};\widetilde{\bbeta}_{k,j})
(\widetilde{\bbeta}_{k,b-1} - \widetilde{\bbeta}_{k,b}) + \bU_{k,b}(\mathcal{D}_{k,b};\widetilde{\bbeta}_{k,b}) = \bm{0},
\end{equation}
where $\widetilde{\bbeta}_{k,1}$ is the MLE at the initial data batch $\mathcal{D}_{k,1}$. Let $\widetilde{\bJ}_{k,b} = \sum_{j=1}^b \bJ_{k,j} (\mathcal{D}_{k,j}; \widetilde{\bbeta}_{k,j} )$ denote the aggregated negative Hessian matrix.  Equation~\eqref{eq:incremental_EE} adds an incremental score function $\bU_{k,b}$ in the updating procedure when the next data batch $\mathcal{D}_{k,b}$ is processed. 

As pointed out in \cite{Luo-Song-2020}, this recursion in the estimation of $(\bbeta, \phi)$ is in a similar spirit to the update of a confidence distribution~\cite{Efron1993}: prior to the consideration of data batch $\mathcal{D}_{k,b}$, the confidence distribution of $\bbeta$ is $\bbeta\mid\mathcal{D}^\star_{k,b-1}\sim \mathcal{N} (\widetilde{\bbeta}_{k,b-1},\text{var}(\widetilde{\bbeta}_{k,b-1}) )$ with $\text{cov}(\widetilde{\bbeta}_{k,b-1})=\widetilde{\phi}_{k,b-1}\widetilde{\bJ}_{k,b-1}^{-1}$, where $\mathcal{D}_{k,b-1}^\star$ is the cumulative dataset up to data batch $\mathcal{D}_{k,b-1}$. We then  update the old confidence distribution, which is treated as a kind of prior distribution, with the distribution of the current data batch $\mathcal{D}_{k,b}$ using the Bayes rule. At the level of the estimating function, this results in the addition of a quantity, $\bU_{k,b}$,  which is the first-order derivative of the log-likelihood function of the current data batch $\mathcal{D}_{k,b}$. 

Solving equation~\eqref{eq:incremental_EE} may be easily done by the following incremental updating algorithm:
\begin{equation}\label{eq:incremental_algorithm}
\widetilde{\bbeta}_{k,b}^{(r+1)} = \widetilde{\bbeta}_{k,b}^{(r)} + 
\left\{\widetilde{\bJ}_{k,b-1} + \bJ_{k,b}(\mathcal{D}_{k,b};\widetilde{\bbeta}_{k,b-1}) \right\}^{-1}
\widetilde{\bU}_{k,b}^{(r)},
\end{equation}
where the adjusted score $\widetilde{\bU}_{k,b}^{(r)} = \widetilde{\bJ}_{k,b-1} (\widetilde{ \bbeta}_{k,b-1} - \widetilde{\bbeta}_{k,b}^{(r)}) + \bU_{k,b}(\mathcal{D}_{k,b};\widetilde{\bbeta}_{k,b}^{(r)})$ is updated over iterations. It is worth pointing out that algorithm~\eqref{eq:incremental_algorithm} uses only subject-level data of the current data batch $\mathcal{D}_{k,b}$ and summary statistics $\{\widetilde{\bbeta}_{k,b-1},\widetilde{\phi}_{k,b-1}, \widetilde{\bJ}_{k,b-1} \}$ from previous data batches. A consistent estimator of the dispersion parameter $\phi$ is updated by a sample size weighted average:
\[
\widetilde{\phi}_{k,b} = \frac{\sum_{j=1}^{b-1}s_{k,j}-p}{\sum_{j=1}^b s_{k,j} - p} \widetilde{\phi}_{k,b-1} + \frac{s_{k,b} - p}{\sum_{j=1}^b s_{k,j} - p} \widehat{\phi}_{k,b},
\]
where $\widehat{\phi}_{k,b}=1/(s_{k,b}) \sum_{i \in \mathcal{D}_{k,b}} \{ (y_i-\widehat{\mu}_i)^2 / v(\widehat{\mu}_i) \}$ is based on the Pearson residuals with $\widehat{\mu}_i=g(\bx_i^\top\widetilde{\bbeta}_{k,b})$ and $v(\cdot)$ is the unit variance function. The negative Hessians for statistical inference on $\bbeta$ are updated incrementally as follows
\begin{equation}\label{eq:incre_J}
\widetilde{\bJ}_{k,b} = \widetilde{\bJ}_{k,b-1} + \bJ_{k,b}(\mathcal{D}_{k,b};\widetilde{\bbeta}_{k,b}).
\end{equation}

Now we have obtained the batch-specific estimators $(\widetilde{\bbeta}_{k,b}, \widetilde{\phi}_{k,b})$, and the batch-specific covariance matrix estimators $\text{var}(\widetilde{\bbeta}_{k,b}) = \widetilde{\phi}_{k,b} \widetilde{\bJ}_{k,b}^{-1}$. Finally, let $\widehat{\bJ}_k=n_k^{-1}\widetilde{\bJ}_{k,Q_k}$, $\widehat{\bbeta}_k=\widetilde{\bbeta}_{k,Q_k}$ and $\widehat{\phi}_k=\widetilde{\phi}_{k,Q_k}$ where $Q_k$ is the index of the last data batch in block $k$.\\
According to \cite{Luo-Song-2020}, under some mild regularity conditions for the data in block $k$ drawn from the homogeneous population $(y_i;\bx_i)\sim f(y;\bx,\bbeta_0,\phi_0)$, $i \in \mathcal{B}_{k}$, the sequentially updated estimator $\widehat{\bbeta}_k=\widetilde{\bbeta}_{k,Q_k}$ is consistent and asymptotically Normally distributed:
\[
\sqrt{n_k}(\widehat{\bbeta}_k - \bbeta_{0}) \overset{d}{\to} \mathcal{N}\left(\bm{0},\phi_{0} \bm{j}^{-1}(\bbeta_{0})\right),  n_k = \sum_{j=1}^{Q_k} s_{k,j} \to\infty,
\]
where $\bm{j}^{-1}(\bbeta_{0})$ is the inverse of the Fisher information for a single observation at the true value $\bbeta_{0}$. The sample counterpart of the Fisher information matrix $\bm{j}(\bbeta_{0})$ is $\widehat{\bJ}_k$; see section \ref{sec:examples} for examples.

\subsection{Combine algorithm}
\label{subsec:alg:combine}
Once each block has been analyzed and block estimators $\widehat{\bbeta}_k$ obtained using the streaming algorithm described in Section \ref{subsec:alg:stream}, we propose an integrated estimator of $\bbeta$ that combines block estimators. We define the PASA estimator for $\bbeta$, which takes the form of a linear update of block estimators $\widehat{\bbeta}_k$:
\begin{align}\label{eq:PASA}
\widehat{\bbeta}_{PASA} &=\left( \sum \limits_{k=1}^K 
n_k \widehat{\phi}^{-1}_k \widehat{\bJ}_k
\right)^{-1} 
\sum \limits_{k=1}^K n_k \widehat{\phi}^{-1}_k \widehat{\bJ}_k \widehat{\bbeta}_k.
\end{align}
Among several methods available in the literature to derive this estimator, we take an approach similar to Efron's confidence distribution \cite{Efron1993} to show the optimality  of the combined estimator (\ref{eq:PASA}); see also \cite{Singh-Xie-Strawderman} and \cite{Xie-Singh}. Under mild regularity conditions, $\sqrt{n_k} \bU_k(\bbeta_0)$ is asymptotically normally distributed with mean $\boldsymbol{0}$ and variance $\phi_0^{-1}\bj(\bbeta_0)$, which is consistently estimated by $\widehat{\phi}_k^{-1} \widehat{\bJ}_k$. Letting $\Phi_p$ be the cumulative distribution function of the $p$-variate standard Normal distribution, the asymptotic confidence estimating function (CEF), as defined in \cite{Hector-Song-JASA}, for block $k$ is $H_k(\bbeta)=\Phi_p ( \sqrt{n_k} \widehat{\phi}^{-1/2}_k \widehat{\bJ}^{1/2}_k \bU_k(\bbeta) )$, and the combined CEF over the $K$ independent blocks is
\begin{align*}
H_C(\bbeta)
&=\prod \limits_{k=1}^K H_k(\bbeta)=\prod \limits_{k=1}^K \Phi_p \left( \sqrt{n_k} \widehat{\phi}^{-1/2}_k \widehat{\bJ}^{1/2}_k  \bU_k(\bbeta) \right) \propto \exp \left\{ -\sum \limits_{k=1}^K n_k \widehat{\phi}^{-1}_k \bU^\top_k(\bbeta) \widehat{\bJ}_k\bU_k(\bbeta) \right\}.
\end{align*}
Maximizing the CEF with respect to $\bbeta$ gives a combined estimator that takes the form
\begin{align}
\widehat{\bbeta}_{C}= \arg \min \limits_{\bbeta} \sum \limits_{k=1}^K n_k \widehat{\phi}^{-1}_k \bU^\top_k(\bbeta) \widehat{\bJ}^{-1}_k \bU_k(\bbeta).
\label{def:GMM-estimator}
\end{align}
The estimator in \eqref{def:GMM-estimator} minimizes a quadratic form of estimating functions and, as such, is a generalized method of moments (GMM) estimator \cite{Hansen}. Under appropriate regularity conditions, $\widehat{\bbeta}_C$ is a consistent estimator of $\bbeta_0$ and asymptotically Normally distributed: $$
\sqrt{N} (\widehat{\bbeta}_C -\bbeta_0) \stackrel{d}{\rightarrow} \mathcal{N}\left(\boldsymbol{0}, {\phi}_0 \bm{j}^{-1}(\bbeta_{0}) \right).
$$ 
The Fisher information matrix $\bm{j}(\bbeta_{0})$ is consistently estimated in a distributed fashion, given by $\sum_{k=1}^K (n_k/N) \widehat\bJ_k$. Because the asymptotic variance is the same as the one in the centralized analysis with the full data processed once, the combined estimator 
$\widehat{\bbeta}_C$ is fully efficient, with no loss of statistical power. Iteratively minimizing the quadratic form in \eqref{def:GMM-estimator}, however, can be computationally intensive, and requires access and repeated loading of the entire dataset, thereby erasing the computational gains afforded by the streaming procedure within each block. Similar to \cite{Hector-Song-JMLR}, we show in the Appendix the following asymptotic identity that gives rise to the definition of the PASA estimator in \eqref{eq:PASA}: 
\begin{align}
\widehat{\bbeta}_C&=\left( \sum \limits_{jk=1}^K n_k \widehat{\phi}^{-1}_k \widehat{\bJ}_k \right)^{-1} \sum \limits_{k=1}^K n_k \widehat{\phi}^{-1}_k \widehat{\bJ}_k \widehat{\bbeta}_k + O_p(N^{-1}) \nonumber \\
 & = \widehat{\bbeta}_{PASA} + O_p(N^{-1}).
\label{eq:GMM-PASA-equiv}
\end{align}
Given the form of the GMM estimator in \eqref{def:GMM-estimator}, the definition of the PASA estimator in \eqref{eq:PASA} follows naturally as a distributed way of linearly updating block estimators $\widehat{\bbeta}_k$. From \eqref{eq:GMM-PASA-equiv} we have that $\widehat{\bbeta}_C-\widehat{\bbeta}_{PASA}=O_p(N^{-1})$, and therefore that $\sqrt{N} (\widehat{\bbeta}_{PASA}  - \bbeta_0) \stackrel{d}{\rightarrow} \mathcal{N}(\boldsymbol{0}, \phi_0\bm{j}^{-1}(\bbeta_{0}) )$, which is again fully efficient. 

This optimal statistical power may be explained as follows.  Due to the asymptotic Normality of block estimators $\{\widehat{\bbeta}_k \}_{k=1}^K$ in the GLM setting, weighting the block estimators by their inverse variance is more efficient than taking an unweighted average of the block estimators. The weighted approach yields the best linear unbiased estimator (BLUE) $\widehat{\bbeta}_{PASA}$, which is asymptotically statistically efficient and communication efficient. It is worth pointing out that both the MapReduce method with only parallel computing, and the online streaming method with only serial computing, are two special cases of the proposed PASA paradigm.

\subsection{Asymptotic properties of the PASA estimator}
\label{subsec:alg:asymptotic}

In this subsection we formalize the inferential properties of the PASA estimator. Suppose $(y_i;\bx_i)$ are $i.i.d.$ samples from a common exponential dispersion model with the underlying true density $f(y;\bx,\bbeta_0,\phi_0)$, $i=1,\dots,N$, where $\bbeta_0$ and $\phi_0$ are the true parameters, respectively.  The mean model of a GLM takes the form $\mu_i=\mathbb{E}(y_i\mid\bx_i)=g(\bx_i^\top\bbeta)$, $\bbeta\in\Theta\subset\mathbb{R}^p$, with variance $V(y_i\mid\bx_i)=\phi v(\mu_i)$, where $\phi>0$ is the dispersion parameter and $v(\cdot)$ is the known unit variance function.  Denote a  consistent estimator of the information matrix using data in $\mathcal{B}_k$ as  $\widehat{\bm{J}}_{k}(\bbeta)=n_k^{-1}\sum_{i\in\mathcal{B}_{k}}\widehat{\bU_i}\widehat{\bU_i}^\top$ where $\widehat{\bU}_i$ is the score function evaluated at the estimated parameter values. If $g$ is the canonical link function, then $\bm{\widehat{J}}_k(\bbeta)=n_k^{-1}\sum_{i\in\mathcal{B}_{k}}\bx_iv(\mu_i)\bx_i^\top$.

Under mild regularity conditions (C1)-(C3) listed in the Appendix, the streaming estimators $\widehat{\bbeta}_k$ were shown to be consistent and asymptotically Normally distributed in \cite{Luo-Song-2020}. They showed that the asymptotic variance of $\widehat{\bbeta}_k$ can be consistently estimated by $\widehat{\phi}_k (n_k\widehat{\bJ}_k)^{-1}$. It then follows from the derivation of $\widehat{\bbeta}_{PASA}$ in the Appendix and from \cite{Hector-Song-JMLR} that, under the additional regularity conditions listed in the Appendix, $\widehat{\bbeta}_{PASA}$ is consistent and asymptotically Normally distributed, that is,
\[
\sqrt{N}(\widehat{\bbeta}_{PASA} - \bbeta_0) \overset{d}{\to}\mathcal{N}\left(\bm{0},\phi_0 \bm{j}^{-1}(\bbeta_{0})\right),
\]
The asymptotic variance is consistently estimated by $(\sum_{k=1}^K n_k \widehat{\phi}_k^{-1} \widehat{\bJ}_k )^{-1}$.  Large-sample properties of MapReduce learning or online streaming learning can be obtained as special cases of the above results. 

\subsection{Computation complexity}
\label{subsec:alg:computation}
A brief summary of the memory and computation complexity of the offline MLE (the gold standard), MapReduce with only parallel computing, and our proposed PASA are shown in Table~\ref{tab:computation} with $N \gg n_k \gg s_{k, b}$ for all $b=1, \ldots, Q_k$, $k=1, \ldots, K$. The running time complexity of the offline MLE is of order $O(Np^2+p^3)$, which is prohibitive when $N$ is large. MapReduce using parallel computing with block size $n_k$ considerably reduces computation complexity by a factor of $1/K$. Our proposed PASA has a similar computation complexity, but it reduces memory complexity by processing each data block in a sequence of $Q_k$ data batches, each of size $s_{k, b}$.
\begin{table}[ht]
\centering
\begin{tabular}{l|c c }
 Method      &Memory &Computation complexity \\
\hline
Offline MLE &$O(Np)$   & $O(Np^2+p^3)$   \\
MapReduce        &$O(n_kp)$ & $O(n_kp^2+p^3)$ \\
PASA        &$O(s_{k, b} p)$ & $O(s_{k, b}Q_kp^2+p^3)$ \\
\end{tabular}
\caption{Computational complexity of the offline MLE, MapReduce and PASA estimators.}
\label{tab:computation}
\end{table}

\section{Examples}
\label{sec:examples}
This section gives details for the implementation of PASA in two GLMs: linear and logistic regression models for Gaussian and binary outcomes respectively.

\subsection{Linear regression}
\label{subsec:examples:linear}
Suppose $y_i\mid\bx_i$ are independently sampled from a Gaussian distribution with mean $\mu_i=\mathbb{E}(y_i\mid\bx_i)=\bx_i^\top\bbeta$ and variance $\text{var}(y_i \mid\bx_i)=\phi$, $i=1, \ldots, N$. We divide the $N$ samples into blocks $\mathcal{B}_k$, $k \in \{1, \ldots, K\}$, and process each block $\mathcal{B}_k$ in a sequence of data batches $\mathcal{D}_{k, b}$, $b=1, \ldots, Q_k$. In batch $\mathcal{D}_{k, b}$, the outcome vector and covariate matrix for all samples can be written as $\by_{k, b}=(y_i)_{i \in \mathcal{D}_{k, b}}$ and $\bX_{k, b}=(\bx_i^\top)_{i \in \mathcal{D}_{k, b}}$ respectively. The score function and the corresponding negative Hessian for data batch $\mathcal{D}_{k,b}$ are, respectively, $\bU_{k,b}(\bbeta)=\bX_{k,b}^\top(\by_{k,b} - \bX_{k,b}\bbeta)$, and $\bJ_{k,b}=\bX_{k,b}^\top\bX_{k,b}, b =1,\ldots, Q_k$, $k=1, \ldots, K$. A closed-form expression for the recursively updated estimator of $\bbeta$ within block $\mathcal{B}_k$ is
\[
\widetilde{\bbeta}_{k,b} = \left(\widetilde{\bJ}_{k,b-1} + \bJ_{k,b} \right)^{-1} \left(\widetilde{\bJ}_{k,b-1}\widetilde{\bbeta}_{k,b-1} + \bX_{k,b}^\top\by_{k,b} \right)
\]
for $b=1,\ldots, Q_k$, with $\widetilde{\bbeta}_0=\bm{0}_p$ and $\widetilde{\bJ}_0=\bm{0}_{p\times p}$ by convention. A consistent estimator of the variance $\phi$ based on $\widetilde{\bbeta}_{k,b}$ is given by the following recursive formula:
\[
\begin{split}
\widetilde{\phi}_{k,b} &= \frac{1}{n_k-p} \sum_{j=1}^b(\by_{k,j}-\bX_{k,j}\widetilde{\bbeta}_{k,b})^\top(\by_{k,j}-\bX_{k,j}\widetilde{\bbeta}_{k,b}) \\
&= \frac{1}{n_k-p} \left\{(n_k-s_b-p)\widetilde{\phi}_{k,b-1} + 
\widetilde{\bbeta}_{k,b-1}^\top \widetilde{\bJ}_{k,b-1} \widetilde{\bbeta}_{k,b-1} + \by_{k,b}^\top\by_{k,b} - \widetilde{\bbeta}_{k,b}^\top\widetilde{\bJ}_{k,b}\widetilde{\bbeta}_{k,b}
\right\},
\end{split}
\]
for $b=1,\ldots, Q_k$. The above $\widetilde{\phi}_{k,b}$ is required for computation of the Fisher information given by $\text{var}(\widetilde{\bbeta}_{k,b})=\widetilde{\phi}_{k,b}(\widetilde{\bJ}_{k,b-1}+\bJ_{k,b})^{-1}$. Note that this estimated variance of $\widetilde{\bbeta}_{k,b}$ gives exactly the same standard error as that of the MLE obtained by fitting the linear model once with the entire data in block $\mathcal{B}_k$. Therefore, the renewable streaming learning carried out within each data block does not lose any statistical efficiency, but is advantageous in data storage and computing speed. 

Finally, we combine the collection of summary statistics $\widehat{\bbeta}_k=\widetilde{\bbeta}_{k,b},\widehat{\bJ}_k=n_k^{-1}\widetilde{\bJ}_{k,b},\widehat{\phi}_k=\widetilde{\phi}_{k,b}$ from each block $\mathcal{B}_k$, $k=1,\dots,K$, to obtain the PASA estimator and its estimated covariance matrix:
\begin{align*}
\widehat{\bbeta}_{PASA}&=\left( \sum \limits_{k=1}^K
n_k \widehat{\phi}^{-1}_k \widehat{\bJ}_k
\right)^{-1} 
\sum \limits_{k=1}^K  n_k\widehat{\phi}^{-1}_k \widehat{\bJ}_k \widehat{\bbeta}_k, \\
\text{cov}(\widehat{\bbeta}_{PASA}) &= \left(\sum_{k=1}^K n_k\widehat{\phi}_k^{-1}\widehat{\bJ}_{k} \right)^{-1}.
\end{align*}

\subsection{Logistic regression}
\label{subsec:examples:logistic}
Suppose $y_i\mid\bx_i$ are independently sampled from a Bernoulli distribution with mean $\mu_i=\mathbb{E}(y_i\mid\bx_i)= \exp(\bx_i^\top \bbeta)/\{1+\exp (\bx_i^\top \bbeta)\}$ and variance $\text{var}(y_i \mid\bx_i)=\phi=1$, $i=1, \ldots, N$. We divide the $N$ samples into blocks $\mathcal{B}_k$, $k \in \{1, \ldots, K\}$ that are divided into batches $\mathcal{D}_{k, b}$, $b=1, \ldots, Q_k$. The score function and the negative Hessian matrix for data batch $\mathcal{D}_{k,b}$, $b=1, \ldots, Q_k$, $k=1, \ldots, K$, are respectively given by
\begin{align*}
\bU_{k,b}(\bbeta) &= \sum_{i\in\mathcal{D}_{k,b}}\bx_{i}\left\{y_{i}-\frac{\exp(\bx_{i}^\top\bbeta)}{1+\exp(\bx_{i}^\top\bbeta)} \right\}, \\
\bJ_{k,b}(\bbeta) &= \sum_{i\in\mathcal{D}_{k,b}} v_{i}\bx_{i}\bx_{i}^\top,
\end{align*}
where $v_{i}(\mu_{i})=\mu_{i}(1-\mu_{i})=\exp(\bx_{i}^\top\bbeta)/\{1+\exp(\bx_{i}^\top\bbeta) \}^2$ is the variance function. The point estimate $\widetilde{\bbeta}_{k,b}$ is updated using the renewable learning in equation~\eqref{eq:incremental_algorithm} and the aggregated observed information matrix $\widetilde{\bJ}_{k,b}$ is updated according to equation~\eqref{eq:incre_J}. Then the PASA estimator and its estimated covariance matrix are
\begin{align*}
\widehat{\bbeta}_{PASA} &=\left( \sum \limits_{k=1}^K n_k\widehat{\bJ}_k
\right)^{-1} 
\sum \limits_{k=1}^K n_k\widehat{\bJ}_k \widehat{\bbeta}_k, \\
\text{cov}(\widehat{\bbeta}_{PASA}) &= \left(\sum_{k=1}^K n_k\widehat{\bJ}_k \right)^{-1}.
\end{align*}

\section{Simulations}
\label{sec:simulations}

\subsection{Setup}
\label{subsec:simulations:setup}
We assess the performance of the proposed PASA estimator through simulation experiments in the linear and logistic regression models. We compare PASA with (i) the offline MLE, the gold standard, obtained by processing the entire data once and (ii) the MapReduce estimator by parallelizing $K$ data blocks. In all simulation experiments, we generate the full dataset with $N$ independent observations from the linear or logistic GLM. This full dataset is randomly partitioned into $K$ data blocks, and data within each block is processed in a sequence of $Q_k\equiv Q$, $k=1, \ldots, K$, data batches. We set $\bbeta_0=(0.2, -0.2, 0.2, -0.2, 0.2)^\top$, the intercept $\bx_{i[1]}\equiv 1$, and $\bx_{i[2:5]}\sim\mathcal{N}_4(\bm{0},\bV_4)$ independently where $\bV_4$ is a $4\times 4$ compound symmetry covariance matrix with correlation $\rho=0.5$.

\subsection{Evaluation of Parameter Estimation}
\label{subsec:simulations:evaluation}
\subsubsection{Different number of blocks $K$ with fixed $Q$}
\label{subsubsec:simulations:evaluation:S1}
\begin{figure}[ht]
\centering
\includegraphics[width=0.6\textwidth]{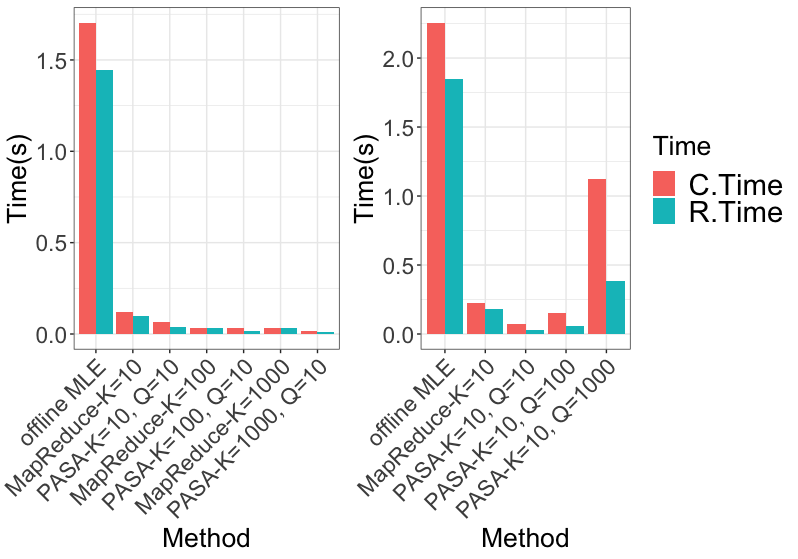}
\caption{Computation time for the linear regression model with $N=10^6$. Left panel shows the computation time at various $K$ and a fixed $Q$ for the offline MLE, MapReduce and PASA methods. Right panel corresponds to the computation time at a fixed $K$ and different $Q$ for the three methods.}
\label{fig:CTime}
\end{figure}

We study the effect of different number of blocks $K$ on point estimation and computation efficiency of the three methods. Note that different $K$ values do not affect the offline MLE since the entire dataset is processed without partitioning, i.e. $K=1$. Tables \ref{tab:linear} and \ref{tab:logistic} report the simulation metrics for the linear and logistic models respectively, over 500 replications. Figure \ref{fig:CTime} shows the computing time for all methods and models.\\
{\bf Bias and coverage probability.} In the linear model, the least squares estimate can be decomposed across different blocks since it is a linear function of the data. Thus, the number of blocks $K$ has minimal effect on estimation bias and coverage probability: both MapReduce and PASA estimators have similar statistical performances to the gold standard offline MLE. In the logistic model, however, larger $K$ leads to a slightly larger bias in both MapReduce and PASA estimators in comparison to the offline MLE. Furthermore, our proposed PASA estimator, which further divides the blocks of data considered in MapReduce into batches, produces exactly the same solution as MapReduce in the linear model, but PASA has a slightly larger bias than MapReduce in the logistic model.\\
{\bf Computation efficiency.} Due to the computational advantages of parallel processing, larger $K$, e.g. $K=100$, improves computation speed by dividing a large dataset into a large number of blocks that are processed simultaneously. In general, as discussed in previous Sections, there is usually a trade-off between computation and statistical efficiency: larger $K$ is associated with slightly larger bias, especially in the logistic model. Our proposed PASA estimator compensates for the computation cost associated with a moderate $K$, e.g. $K=10$, by processing each data block in a sequence of $Q$ data batches. This yields computation times for PASA that are twice as fast as the computation time for MapReduce. Our simulation results demonstrate that with moderate $K$ and $Q$, the statistical performance with PASA is very close to the statistical performance of the offline MLE, but that the computation time of PASA is only 1/14 and 1/38 of the computation time of the offline MLE in the linear and logistic models respectively.\\
\begin{table*}[ht]
\centering
\resizebox{\textwidth}{!}{
\begin{tabular}{l | c | c c c c | c c c c| c c c c}
\multicolumn{14}{c}{\bf Linear Model} \\
$N=10^5$  &Offline MLE &\multicolumn{4}{c|}{MapReduce} &\multicolumn{4}{c|}{PASA, fixed $Q$} &\multicolumn{4}{c}{PASA, fixed $K$}\\
$K$      &1     &1  &10  &100  &1000    &1  &10  &100  &1000  &\multicolumn{4}{c}{10}\\
$Q$      &1    &\multicolumn{4}{c|}{1} &\multicolumn{4}{c|}{10} &1  &10  &100  &1000\\
\hline
A.bias$\times 10^{-3}$ &3.067 &3.067 &3.067 &3.070 &3.101 &3.067 &3.067 &3.070 &3.101 
&3.067 &3.067 &3.067 &3.067\\
ASE$\times 10^{-3}$ &3.833 &3.833 &3.832 &3.829 &3.792 &3.833 &3.832 &3.829 &3.792 
&3.832 &3.832 &3.832 &3.832\\
ESE$\times 10^{-3}$ &3.822 &3.822 &3.823 &3.827 &3.859 &3.822 &3.823 &3.827 &3.859
&3.823 &3.823 &3.823 &3.823\\
CP &0.950 &0.950 &0.950 &0.948 &0.947 &0.950 &0.950 &0.948 &0.947
&0.950 &0.950 &0.950 &0.950\\
C.Time(s) &0.278 &0.205 &0.038 &0.012 &0.010 &0.082 &0.019 &0.016 &0.024 
&0.010 &0.021 &0.060 &0.408\\
R.Time(s) &0.236 &0.166 &0.030 &0.011 &0.010 &0.041 &0.012 &0.011 &0.012
&0.005 &0.014 &0.033 &0.184\\
\end{tabular}
}
\caption{Simulation results under the linear model summarized over 500 replications, with fixed $N=10^5$, $p=5$. ``A.bias", ``ASE", ``ESE" and ``CP" represent the mean absolute bias, the averaged estimated standard error of the estimates, the empirical standard error, and the coverage probability, respectively. ``C.Time" and ``R.Time" respectively denote computation time and running time in seconds.}
\label{tab:linear}
\end{table*}

\begin{table*}[ht]
\centering
\resizebox{\textwidth}{!}{
\begin{tabular}{l|c | c c c  | c c c | c c c}
\multicolumn{11}{c}{\bf Logistic Model} \\
$N=10^5$  &Offline MLE &\multicolumn{3}{c|}{MapReduce} &\multicolumn{3}{c|}{PASA, fixed $Q$} &\multicolumn{3}{c}{PASA, fixed $K$}\\
$K$      &1     &1  &10  &100     &1  &10  &100  &\multicolumn{3}{c}{10}\\
$Q$      &1    &\multicolumn{3}{c|}{1} &\multicolumn{3}{c|}{10} &1 &10 &100\\
\hline
A.bias$\times 10^{-3}$ &6.251 &6.251 &6.251 &6.311  &6.251 &6.256 &6.844 
&6.251 &6.256 &6.262\\
ASE$\times 10^{-3}$ &7.817 &7.817 &7.819 &7.842  &7.817 &7.823 &7.885 
&7.819 &7.823 &7.827\\
ESE$\times 10^{-3}$ &7.831 &7.831 &7.827 &7.778  &7.831 &7.819 &7.692
&7.826 &7.819 &7.809\\
CP &0.952 &0.952 &0.952 &0.952  &0.952 &0.953 &0.935 
&0.952 &0.953 &0.955\\
C.Time(s) &0.531 &0.469 &0.016 &0.013  &0.094 &0.013 &0.016 
&0.015 &0.022 &0.054\\
R.Time(s) &0.493 &0.428 &0.010 &0.012  &0.058 &0.008 &0.010 
&0.011 &0.015 &0.025\\
\end{tabular}
}
\caption{Simulation results for the logistic model are summarized over 500 replications, with fixed $N=10^5$, $p=5$. ``A.bias", ``ASE", ``ESE" and ``CP" represent the mean absolute bias, the averaged estimated standard error of the estimates, the empirical standard error, and the coverage probability, respectively. ``C.Time" and ``R.Time" respectively denote computation time and running time in seconds.}
\label{tab:logistic}
\end{table*}

\subsubsection{Fixed $K$ but varying number of data batches $Q$}
\label{subsubsec:simulations:evaluation:S2}
We study the effect of different number of batches $Q$ on point estimation and computation efficiency of the three methods with a fixed number of blocks $K=10$. Tables \ref{tab:linear} and \ref{tab:logistic} report the simulation metrics for the linear and logistic models respectively, over 500 simulations. Figure \ref{fig:CTime} shows the computing time for all methods and models. \\
\textbf{Bias and coverage probability.} Increasing the number of data batches $Q$ with a fixed $K$ does not affect estimation bias or coverage probability in the linear model, but the bias becomes slightly larger in the logistic regression model. This loss of statistical efficiency is due to approximations in the sequential estimation and inference for the logistic model and is exacerbated with larger $Q$.\\
\textbf{Computation efficiency.} Under a fixed $K$, computation time for PASA increases with $Q$ because the computational cost of matrix inversions increases linearly with $Q$. At each iteration, the computation complexity of PASA and MapReduce are, respectively, $O(s_{k,b}p^2+p^3)$ and $O(n_kp^2+p^3)$, with $n_k \gg s_{k,b}$. Therefore, PASA is faster than MapReduce as long as $Q$ is not too large. PASA becomes more advantageous with larger $N$, say $N=10^6$; see the right panel in Figure~\ref{fig:CTime}.  

\section{Prediction Model Building Application}
\label{sec:application}

We illustrate the application of our proposed PASA estimator by building a prediction model for conversion outcomes in online search advertising. Prediction model building is notoriously time consuming since it proceeds by iteratively refitting models. PASA significantly reduces the computational burden associated with this task. The advertising company Criteo released a publically available dataset recording 90 days of Criteo traffic data \cite{Tallis-etal}. Observations are clicks on a product related advertisement performed by individuals who expressed interest by an online search. The recorded data include the product's target age group and gender, the type of device used for the click, the number of clicks on the product related advertisement received in the last week, and whether the click led to a purchase of the product, termed conversion, within a 30 day window after the click. 

\subsubsection{Logistic model training}
After cleaning and preprocessing, we model the probability of a conversion using $N=3,367,000$ observations through the training of a logistic regression model. Covariates consist of standardized number of clicks ($clicks$), one dummy indicator variable for the second age group ($age_2$), two dummy indicator variables for the second and third device type groups ($device_2$, $device_3$), and two dummy indicator variables for the second and third gender groups ($gender_2$, $gender_3$).

\begin{figure}[ht]
\centering
\includegraphics[width=0.6\textwidth]{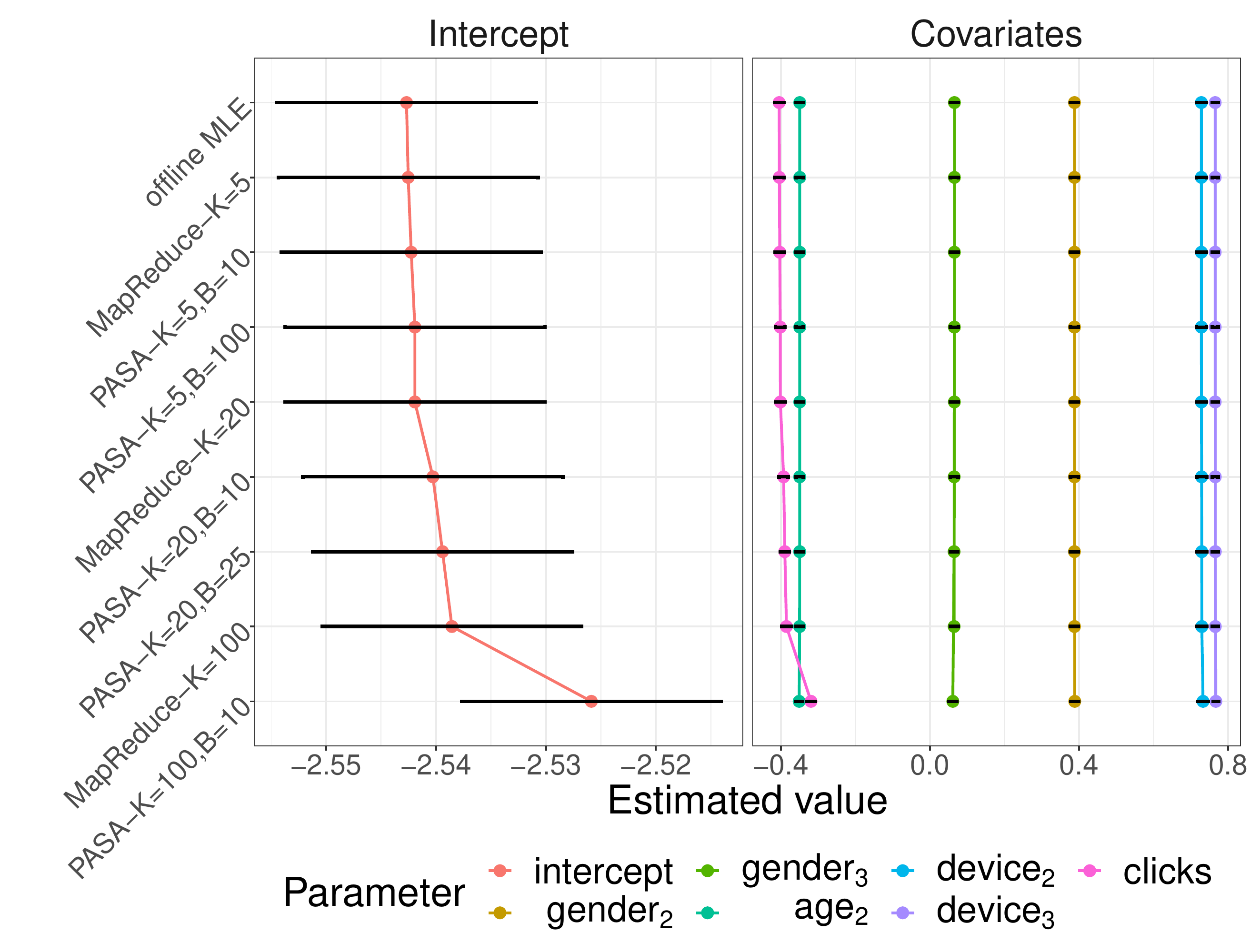}
\caption{Estimates of the base model parameters for the offline MLE, MapReduce and PASA in five settings. Error bars around estimates correspond to estimated standard deviation. \label{data:estimates}}
\end{figure}

We first build a base model with the six covariates and an intercept to compare the performance of the offline MLE, the MapReduce estimator with $K\in \{5, 20, 100\}$, and PASA in five settings: (i) $K=5$ and $Q_k=10$, (ii) $K=5$ and $Q_k=100$, (iii) $K=20$ and $Q_k=10$, (iv) $K=20$ and $Q_k=25$, and (v) $K=100$ and $Q_k=10$. In each setting, $Q_k\equiv Q$ for $k=1, \ldots, K$. Parameter estimates and estimated standard errors are plotted in Figure \ref{data:estimates}. C.Time and R.Time are given in Table \ref{data:times}. It is not surprising that, with such a large sample size, all covariates are statistically significant in all models. From Figure \ref{data:estimates}, the PASA estimator deteriorates for $K=100$, as is expected from simulation take-aways. For $K<100$, the PASA estimator achieves similar estimates and confidence intervals to the MapReduce and offline MLE estimators, reinforcing our confidence in its performance. From Table \ref{data:times}, PASA outperforms MapReduce and offline MLE estimators in C.Time and R.Time in all settings except $K=100$ and $K=20,Q=25$. We identify the optimal PASA estimator in terms of statistical and computational performance as the estimator in setting $K=20,Q=10$. The computational burdens of the offline MLE are particularly evident, with C.time of 20.97s and R.Time of 15.20s. This is 8 and 11 times slower, respectively, than the slowest PASA C.Time of 2.69s and R.Time of 1.38s in case (i) and 22 and 31 times slower, respectively, than the fastest PASA estimator C.Time of 0.95s and R.Time of 0.49s respectively.
\begin{table}[ht]
\centering
\begin{subtable}[ht]{0.49\textwidth}
\centering
\subfloat[]{
\begin{tabular}{l | c}
 \multicolumn{2}{c}{offline MLE} \\
 \multicolumn{1}{r|}{$K$} & $1$ \\
 \multicolumn{1}{r|}{$Q$} & $1$ \\ 
  \hline
C.Times(s) & 20.97 \\ 
R.Time(s) & 15.20 \\ 
\end{tabular}
}
\quad
\subfloat[]{
\begin{tabular}{l | cccc}
 \multicolumn{4}{c}{MapReduce} \\
 \multicolumn{1}{r|}{$K$}& $5$ & $20$ & $100$ \\ 
\multicolumn{1}{r|}{$Q$} & \multicolumn{3}{c}{$1$} \\
  \hline
C.Times(s) & 3.38 & 1.04 & 0.21 \\ 
R.Time(s) & 2.34 & 0.73 & 0.13 \\ 
\end{tabular}
}
\end{subtable}
\hfill
\begin{subtable}[ht]{0.49\textwidth}
\centering
\subfloat[]{
\centering\begin{tabular}{l | ccccc}
\multicolumn{6}{c}{PASA} \\
\multicolumn{1}{r|}{$K$} & \multicolumn{2}{c}{$5$} & \multicolumn{2}{c}{$20$} & $100$ \\ 
\multicolumn{1}{r|}{$Q$} & $10$ & $100$ & $10$ & $25$ & $10$ \\
  \hline
C.Times(s) & 2.69 & 2.01 & 0.95 & 1.19 & 0.59 \\ 
R.Time(s) & 1.38 & 0.52 & 0.49 & 0.82 & 0.51 \\ 
\end{tabular}
}
\end{subtable}
\caption{Computing time for estimation of $\bbeta$ in the base model for the Criteo data conversion model using (a) offline MLE, (b) MapReduce and (c) PASA methods. ``C.Time'' and ``R.Time'' respectively denote computation time and running time in seconds. \label{data:times}}
\end{table}

\subsubsection{Prediction model building} 
We highlight the computational gains of PASA when iterated over multiple model building and evaluation steps using a prediction model building example. Using the PASA estimator in setting $K=20,Q=10$, we build a prediction model using forward-selection based on the base model. The forward-selection proceeds as follows. At each forward step, we add one term to the model from the previous step and fit this model on training data (the first $15$ blocks of the data) and evaluate the AUC on the testing data (the last $5$ blocks); the term leading to the highest AUC is added permanently to the model. Terms under consideration at each step are all interaction terms between all variables in the base model. Note that while building the prediction model, after adding the  $device_3 clicks$ covariate we remove the $gender_2 device_3$ covariate from consideration due to poor identification. The final selected prediction model is the model with the highest AUC, which consists of the following main and interaction covariates: $gender_2$, $gender_3$, $age_2$, $device_2$, $device_3$, $clicks$, $age_2  clicks$, $gender_3  age_2$, $gender_3  age_2  device_2$, $gender_3  clicks$, $device_2  click$, $gender_3  device_2  clicks$, $gender_3  age_2  clicks$. The forward prediction procedure with PASA evaluates 435 models in C.Time 1.36 hours and R.Time 58 minutes.

The resulting prediction model has AUC 0.6440, an improvement over the base model AUC of 0.6398. This apparently small difference should be interpreted with caution. To elucidate this, in Table \ref{table-corrected}, using the testing data, we compare the number of false negatives (FN), false positives (FP) and false (F) predictions of the base and prediction models. We report the number of predictions that were incorrect in the base model and correct in the prediction model (CORR). Let $\mathbbm{1}(\cdot)$ denote the indicator function. Predicting the conversion outcome based on the cutoff $\mathbbm{1}(\widehat{p}_i>0.1)$ of the predicted probability $\widehat{p}_i$ for observation $i$, the prediction model with the interactions corrects 142,337 falsely predicted outcomes in the base model. Based on the cutoff $\mathbbm{1}(\widehat{p}_i>0.05)$, the prediction model corrects 214,145 falsely predicted outcomes of the base model. These corrections occur primarily by correcting false positives in the base model; in other words, the prediction model more accurately predicts a conversion. Advertising companies could better target their advertisements with this improved knowledge of users who will eventually purchase a product. It is also clear that the computational advantages of PASA extend to more complex data settings.
\begin{table}[ht]
\centering
\begin{tabular}{ll | cccc}
$\widehat{p}_i$ cutoff & model & FN & FP & F & CORR \\ 
\hline
\multirow{2}{*}{$\widehat{y}_i=\mathbbm{1}(\widehat{p}_i>0.1)$} & base & 34281 & 286358 & 320639 & \multirow{2}{*}{142337} \\ 
 & prediction & 54248 & 181107 & 235355 & \\ 
 \multirow{2}{*}{$\widehat{y}_i=\mathbbm{1}(\widehat{p}_i>0.05)$} & base & 1024 & 719820 & 720844 & \multirow{2}{*}{214145} \\ 
 & prediction & 17783 & 517256 & 535039 & \\ 
\end{tabular}
\caption{FN, FP, F and CORR predictions for prediction and base models using different cutoffs of predicted probability to predict the conversion outcome. \label{table-corrected}}
\end{table}

\section{Discussion}
\label{sec:discussion}

The proposed PASA delivers a new accelerated algorithm to perform fast, scalable and efficient statistical analyses of very large datasets. The main mechanism of this powerful doubly distributed paradigm is a unique combination of parallel and streaming algorithms that leverages recent developments in high-speed computing infrastructure such as the MapReduce paradigm and the Rho architecture. The iterated division of data yields small data batches that can be loaded and analysed quickly. Simulations and the prediction model building application highlight the tremendous computational gains of the proposed PASA approach and confirm that the PASA estimator loses little  estimation efficiency in comparison to the gold standard centralized data analysis and popular MapReduce parallelized analysis.

In this paper, PASA is implemented for estimation and inference in GLM, a class of models that encompasses a broad scope of valuable and preferred data analysis tools. The asymptotic properties ensure that the PASA estimator is estimation consistent and asymptotically Normal. Generalization of the PASA paradigm to handle other data types, such as time-to-event and longitudinal outcomes, is of interest and worth further exploration. These generalizations require some efforts to generalize the theoretical foundation of Efron's confidence distribution \cite{Efron1993} and Hansen's generalized method of moments \cite{Hansen}. 

In practice, the choice of the number $K$ of data blocks and the number $Q_k$ of data batches is key to the trade-off between speed and efficiency, as illustrated in the simulations. Larger $K$ generally yields accelerated computation time but poorer statistical performance, whereas smaller and larger $Q_k$ yields improved statistical performance but slower computation time. As a general rule of thumb, we recommend specifying smaller $K$ and larger $Q_k$, although the particulars of each data analysis should be considered prior to implementation of PASA.

We anticipate PASA will be useful for a variety of data analytic goals, such as modelling, prediction and classification. The illustration of PASA for prediction model building underscores the feasibility of complex statistical tasks with very large data. Possible applications include public health, marketing, economics, forecasting, and many more.

\section*{Acknowledgment}

The third author (Song)'s research was partially funded by NSF DMS\#1811734.  

\bibliographystyle{apalike}
\bibliography{bibliography-20200623}

\begin{thebibliography}{}

\bibitem[Efron, 1993]{Efron1993}
Efron, B. (1993).
\newblock Bayes and likelihood calculations from confidence intervals.
\newblock {\em Biometrika}, 80:3--26.

\bibitem[Glass, 1976]{Glass}
Glass, G.~V. (1976).
\newblock Primary, secondary, and meta-analysis of research.
\newblock {\em Educational Researcher}, 5(10):3--8.

\bibitem[Hansen, 1982]{Hansen}
Hansen, L.~P. (1982).
\newblock Large sample properties of generalized method of moments estimators.
\newblock {\em Econometrica}, 50(4):1029--1054.

\bibitem[Hector and Song, 2020a]{Hector-Song-JASA}
Hector, E.~C. and Song, P. X.-K. (2020a).
\newblock A distributed and integrated method of moments for high-dimensional
  correlated data analysis.
\newblock {\em Journal of the American Statistical Association},
  \textnormal{DOI: 10.1080/01621459.2020.1736082}:1--14.

\bibitem[Hector and Song, 2020b]{Hector-Song-JMLR}
Hector, E.~C. and Song, P. X.-K. (2020b).
\newblock Doubly distributed supervised learning and inference with
  high-dimensional correlated outcomes.
\newblock {\em Journal of Machine Learning Research}, 21:1--35.

\bibitem[Jordan, 2013]{Jordan}
Jordan, M.~I. (2013).
\newblock On statistics, computation and scalability.
\newblock {\em Bernoulli}, 19(4):1378--1390.

\bibitem[J{\o}rgensen, 1997]{Jorgensen1997}
J{\o}rgensen, B. (1997).
\newblock {\em The theory of dispersion models}.
\newblock Chapman and Hall, London.

\bibitem[Luo and Song, 2020]{Luo-Song-2020}
Luo, L. and Song, P. X.-K. (2020).
\newblock Renewable estimation and incremental inference in generalized linear
  models with streaming datasets.
\newblock {\em Journal of the Royal Statistical Society: Series B}, 82:69--97.

\bibitem[Robbins and Monro, 1951]{Robbins1951}
Robbins, H. and Monro, S. (1951).
\newblock A stochastic approximation method.
\newblock {\em The Annals of Mathematical Statistics}, 22(3):400--407.

\bibitem[Sakrison, 1965]{Sakrison1965}
Sakrison, D.~J. (1965).
\newblock Efficient recursive estimation: application to estimating the
  parameter of a covariance function.
\newblock {\em International journal of engineering science}, 3(4):461--483.

\bibitem[Singh et~al., 2005]{Singh-Xie-Strawderman}
Singh, K., Xie, M., and Strawderman, W.~E. (2005).
\newblock Combining information from independent sources through confidence
  distributions.
\newblock {\em The Annals of Statistics}, 33(1):159--183.

\bibitem[Song, 2007]{Song}
Song, P. X.-K. (2007).
\newblock {\em Correlated Data Analysis: Modeling, Analytics, and
  Applications}.
\newblock Springer Series in Statistics.

\bibitem[Tallis and Yadav, 2018]{Tallis-etal}
Tallis, M. and Yadav, P. (2018).
\newblock Reacting to variations in product demand: An application for
  conversion rate (cr) prediction in sponsored search.
\newblock {\em arXiv preprint arXiv:1806.08211}.

\bibitem[Tang and Song, 2016]{Tang-Song}
Tang, L. and Song, P. X.-K. (2016).
\newblock Fused lasso approach in regression coefficients clustering --
  learning parameter heterogeneity in data integration.
\newblock {\em Journal of Machine Learning Research}, 17:1--23.

\bibitem[Toulis and Airoldi, 2015]{Toulis2015MLE}
Toulis, P. and Airoldi, E.~M. (2015).
\newblock Scalable estimation strategies based on stochastic approximations:
  classical results and new insights.
\newblock {\em Statistics and computing}, 25(4):781--795.

\bibitem[Wang et~al., 2012]{Wang-Wang-Song-2012}
Wang, F., Wang, L., and Song, P. X.~K. (2012).
\newblock Quadratic inference function approach to merging longitudinal
  studies: validation and joint estimation.
\newblock {\em Biometrika}, 99(3):755--762.

\bibitem[Xie and Singh, 2013]{Xie-Singh}
Xie, M. and Singh, K. (2013).
\newblock Confidence distribution, the frequentist distribution estimator of a
  parameter: a review.
\newblock {\em International Statistical Review}, 81(1):3--39.

\bibitem[Xie et~al., 2011]{Xie-Singh-Strawderman}
Xie, M., Singh, K., and Strawderman, W.~E. (2011).
\newblock Confidence distributions and a unifying framework for meta-analysis.
\newblock {\em Journal of the American Statistical Association},
  106(493):320--333.

\bibitem[Zellner, 1962]{Zellner}
Zellner, A. (1962).
\newblock An efficient method of estimating seemingly unrelated regressions and
  tests for aggregation bias.
\newblock {\em Journal of the American Statistical Association},
  57(298):348--368.

\end{thebibliography}

\end{document}